# A Construction-Phase Digital Twin Framework for Quality Assurance and Decision Support in Civil Infrastructure Projects

Md Asiful Islam[1], Shanto Jouerder[2], Md Sabit As Sami[3], Afia Jahin Prema[4]

**Corresponding Author:** Md Asiful Islam (Email: mxi2776@mavs.uta.edu, asiful.islam.me@gmail.com)

[1] Intertek PSI, Dallas, TX 75243, USA (e-mail: asiful.islam.me@gmail.com, mdasiful.islam@intertek.com).

[2] Bangladesh University of Engineering and Technology, Dhaka, Bangladesh (e-mail: shanto.jouerder@gmail.com).

[3] Pabna University of Science and Technology, Pabna 6600, Bangladesh (e-mail: sabitassami@gmail.com).

[4] University of Texas at Arlington, Arlington, TX 76010, USA (email: axp9355@mavs.uta.edu)

**Abstract**

Quality assurance (QA) during construction often relies on inspection records and laboratory test results that become available days or weeks after work is completed. On large highway and bridge projects, this delay limits early intervention and increases the risk of rework, schedule impacts, and fragmented documentation. This study presents a construction-phase digital twin framework designed to support element-level QA and readiness-based decision making during active construction. The framework links inspection records, material production and placement data, early-age sensing, and predictive strength models to individual construction elements. By integrating these data streams, the system represents the evolving quality state of each element and supports structured release or hold decisions before standard-age test results are available. The approach does not replace established inspection and testing procedures. Instead, it supplements existing workflows by improving traceability and enabling earlier, data-informed quality assessments. Practical considerations related to data integration, contractual constraints, and implementation challenges are also discussed. The proposed framework provides a structured pathway for transitioning construction QA from delayed, document-driven review toward proactive, element-level decision support during construction.

**Keywords:** Construction-phase digital twin; Digital twin in Quality assurance; Early-age sensing; Predictive analytics; Data integration in construction.

## 1 Introduction

The construction industry is increasingly adopting digital technologies to manage the complexity of large infrastructure projects and the growing volume of heterogeneous information generated during planning, construction, and operation. Among these technologies, the digital twin has emerged as a data-connected representation of a physical asset or process that is continuously informed by field observations. Grieves and Vickers formalized the concept as a dynamic linkage between the physical system, its digital representation, and the data connections that enable continuous information exchange and analysis [1]. With advances in sensing, data management, and computation, digital twin concepts have rapidly expanded from their early roots in manufacturing to broader engineering domains.

Digital twin research has developed beyond industrial systems to include buildings, transportation infrastructure, and urban environments. Prior studies show that digital twins can support visualization, integration of fragmented information, and lifecycle-oriented analysis by linking geometric models with inspection records, sensor measurements, and analytical tools [2–7]. Bridge, railway, and large complex buildings infrastructure-oriented applications have proved useful in tracking and assessing the condition of infrastructures and their performance [4,7,8]. A digital twin framework based on parametric BIM, real-time sensor data, energy simulation, and machine learning was proposed to assist greenhouse gas emission reduction in smart campus buildings [9]. These initiatives underscore the importance of digital twins as integrative environments that have the potential to bring about fragmented project information to a coherent digital form.

Recent work has also focused on the application of digital twins during the construction phase. Applications have been reported in progress monitoring, visualization of construction sequences by BIM, combining BIM and field data, and in reality, capture technologies, UAV photogrammetry and laser scanning to update digital models [10]. Reviews and conceptual studies also elaborate on new developments in construction digital twins, such as semantic modeling, data interoperability, and challenges to adoption among project stakeholders [11–13]. Collectively, this literature indicates growing interest in applying digital twin concepts earlier in the project lifecycle, rather than limiting them to post-construction asset management.

Despite these advances, existing construction digital twin implementations are still limited to mainly information aggregation, monitoring and visualization. Conceptual differences between static models and fully realized digital twins highlight that many of the reported systems do not have continuous data integration and operational decision logic [11,13–15]. As a result, digital twins may enhance situational awareness on site, but their ability to support systematic and decision-oriented workflows for construction quality assurance remains limited.

Motivated by these limitations, this paper presents a construction-phase digital twin framework designed to support quality-related decision making during active construction. The proposed approach integrates inspection records, material production and placement information, and project geometry in a common, element-centric digital environment. Rather than replacing established inspection and testing practices, the framework is designed to supplement current workflows by enabling traceable, readiness-based QA assessment as construction progresses.

Recent studies have explored the use of digital twins in construction, including applications in data integration, progress tracking, inspection support, and quality management. As shown in Table 1, most of these frameworks aim to improve visibility, coordination, and information sharing during construction. Some studies use quality-related data such as inspection records or defect reports, but this information is usually applied for monitoring and documentation rather than for making acceptance decisions. Even when construction quality is discussed directly, quality assurance is often treated at a managerial level, without clearly explaining how inspection results and material performance lead to release or hold decisions for the next construction step.

A common limitation is the absence of element-level QA logic. Quality is rarely modeled as a time-dependent condition tied to specific construction elements, and clear readiness or acceptance criteria are not typically built into the digital twin system. As a result, quality assurance remains largely reactive, relying on delayed test results and manual coordination outside the digital environment. These limitations highlight the need for a construction-phase digital twin that goes beyond visualization and monitoring and instead supports clear, traceable, and decision-based QA during active construction. The present work addresses this need by proposing an element-centric digital twin model that integrates inspection data and material behavior to support structured QA decisions.

**Table 1.** Review of digital twin–based approaches for construction quality assurance

| Authors (Ref) | DT Focus | Construction Phase & QA | Key Gap |
|---|---|---|---|
| Boje et al. (2020) [2] | Semantic construction digital twin | Construction phase considered; QA treated as data integration | Semantic integration achieved; no rule-based QA decision mechanism |
| Deng et al. (2021) [3] | DT-driven smart construction | Construction phase discussed; QA framed at management level | QA not operationalized at element level |
| Lu et al. (2020) [4] | Infrastructure digital twin | Primarily post-construction; QA implicit via asset condition | No explicit construction-phase QA workflow |
| Tao et al. (2019) [5] | Digital twin state-of-the-art | Lifecycle overview; QA not construction-specific | Construction QA not addressed |
| Wright and Davidson (2020) [11] | Model vs digital twin distinction | Conceptual framing applied to construction | No construction-phase, decision-oriented QA implementation |
| Opoku et al. (2021) [15] | Systematic review of DT in construction | Construction phase included; QA discussed descriptively | Lack of decision-oriented QA use cases |
| Ghansah & Edwards (2024) [16] | Digital technologies for construction QA | Construction QA reviewed; DT treated as enabling concept | No DT-based QA decision framework |
| Liu et al. (2023) [17] | DT-based quality control for steel structures | Construction QC via compliance checks | QA limited to static checks, no sequencing logic |

## 2 Construction-Phase Quality Assurance Framework

The suggested framework presents the QA in the construction phase as a realistic, data-driven workflow, which mirrors the way in the field construction activities are being conducted. Instead of considering quality as a set of unrelated inspection reports and laboratory tests, the framework brings quality data to the construction elements and assesses their preparedness as the construction progresses. This method has been developed as a response to the constraints perceived in previous construction digital twin research, where data on quality is frequently combined to be visualized or reported, but not to inform an express acceptance or release decision throughout construction [2,3,15]. Fig. 1 shows the workflow integrating inspections and materials testing for QA.

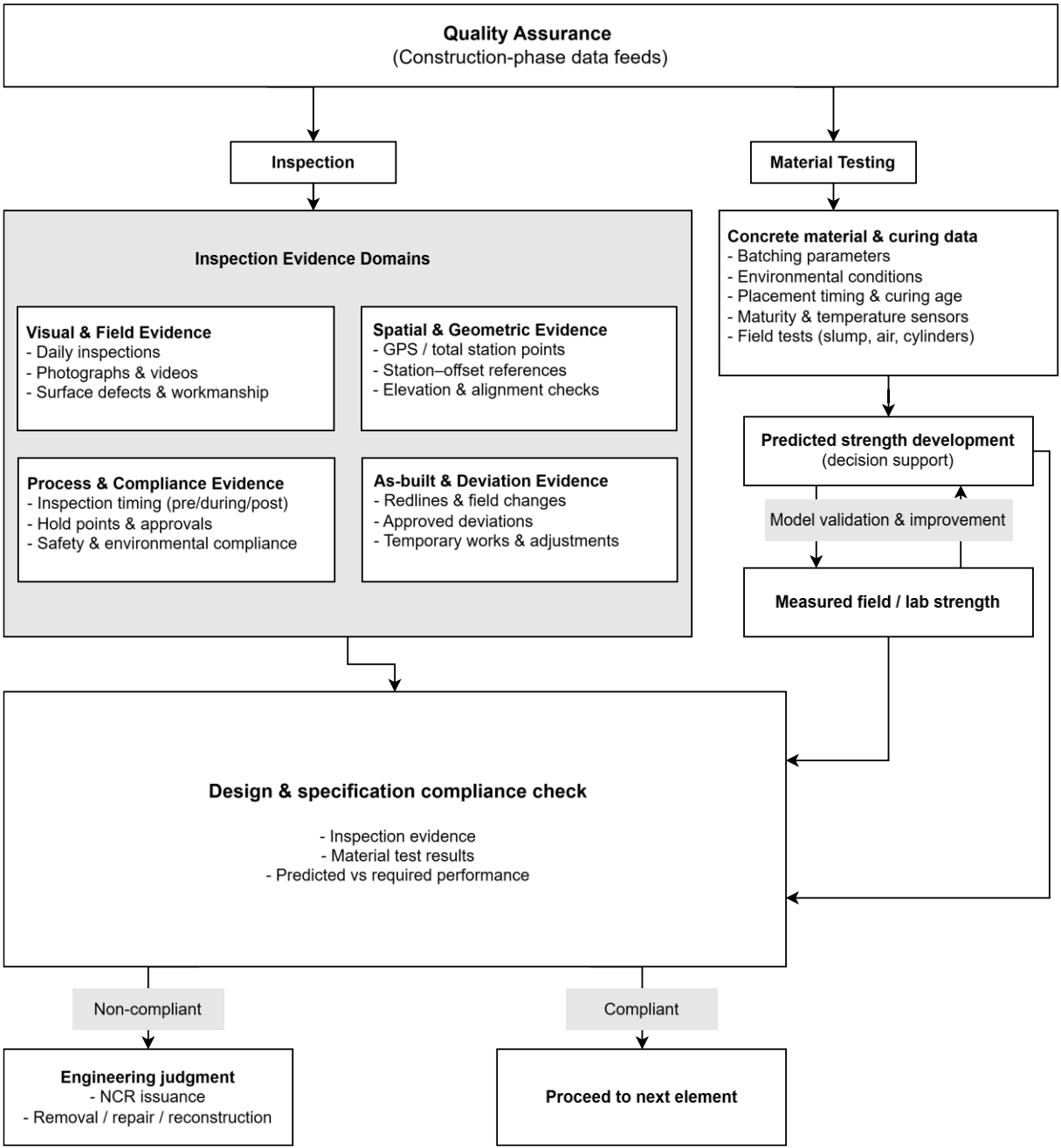

Fig. 1. Construction-phase QA workflow integrating inspections and materials testing.

## 2.1 Inspection-Side Data Integration

On the inspection side, the framework connects to information that is already routinely collected during construction. This includes daily inspection reports, pre-placement and post-placement checklists, inspector comments, photographic records, and deficiency or punch list items. Instead of storing these records as isolated documents, the framework links them directly to individual construction elements.

Each inspection record is mapped using spatial references such as station-offset, GPS coordinates, or survey control points. This allows inspection outcomes to be evaluated at the element level rather than at the document level. While previous digital twin studies have discussed data integration and construction monitoring [2,10], they have not explicitly tied inspection information to quality decisions for specific elements.

Geospatial and visual inputs, including total station measurements, UAV imagery, and orthomosaic products, are incorporated to provide context on geometry, alignment, and construction progress. These inputs are used primarily to support documentation and situational awareness rather than automated vision-based interpretation. This approach is consistent with observations that many construction digital twins focus on monitoring and visualization functions [11].

The framework also considers the timing of inspections relative to construction activities. Pre-placement inspections, placement windows, curing periods, and formal hold points are tracked to determine whether required inspections have been completed within the correct sequence. This ensures that inspection completeness is assessed in relation to the actual construction timeline.

Only inspection observations that directly affect construction quality are used in QA decision logic. Safety or environmental observations are incorporated only when they influence placement conditions, structural stability, or curing performance. Approved field changes, as-built adjustments, and documented deviations are also included so that quality evaluation reflects actual construction conditions rather than design intent alone. This addresses a limitation in prior construction QA research, where analysis often remains document-based and disconnected from field reality [16].

## 2.2 Material Testing and Early-Age Behavior

Simultaneously, the material testing element goes beyond the traditional acceptance testing as it explicitly considers the behavior of materials throughout the construction. In the case of concrete, the batching parameters, delivery and placement records, the environmental conditions, the duration of curing, and data of embedded temperature or maturity sensors are used in combination to explain how the development of the concrete strength takes place in place over time. This represents known results that the strength development of concrete is not only related to the proportions of the mixtures, it is also related to temperature history and curing conditions, which can be only described in part through laboratory-cured specimens [18–20].

The framework includes construction-stage variables to determine the strength development as a time-dependent, continuous process instead of only using discrete test outcomes at different ages. This method also agrees with the principles of maturity-based strength estimation formalized in ASTM C1074 [21] that allow in-place strength to be estimated using embedded sensors depending on temperature history. The experimental research has shown that maturity-based approaches can offer meaningful estimates of the strength development in the field conditions in case of adequate calibration.

Early-age strength development is estimated using predictive models to help in estimating development before the standard-age test results are made available. These predictions are not meant to substitute physical testing, but to complement them by giving early warnings of whether or not placed material is heading to compliance or a possible deficiency. Validation anchors are laboratory-cured or field-cured test results becoming available and make it possible to detect discrepancies between predicted and measured strengths and further improve predictive performance over time. This predictive-validation feedback process is based on recent studies that show that data-driven concrete strength prediction through variables of the construction phase is possible at the industry scale [22].

### 2.3   Element-Level QA Decisions

The central feature of the framework is the integration of inspection data and material performance to support element-level quality decisions. For each construction element, inspection completeness and material behavior are evaluated against design and specification requirements. Based on this assessment, the element is assigned a clear QA state such as pending, released, held, or non-conforming.

When inspection requirements are satisfied and material performance trends toward compliance, the element may proceed to the next construction stage with documented justification. If predicted or measured results indicate potential non-compliance, the system supports early engineering review and corrective actions such as extended curing, additional testing, process adjustments, or removal and replacement.

By treating quality as a condition that evolves over time rather than a final check after testing, the framework enables earlier intervention and clearer traceability between evidence and decisions.

## 3   System Architecture for Construction-Phase QA Digital Twin

### 3.1   Overview of the System Architecture

The presented digital twin construction-phase QA consists of a layered system architecture that will be used to process heterogeneous construction data and integrate it into a single entity-centric format, and provide controlled QA decision-making in on-site construction. The architecture is structured into five logical layers as illustrated in Fig. 2 and Fig. 3:

(1) Construction-phase data sources
(2) Data ingestion and harmonization
(3) The element-centric QA digital twin core
(4) Analytics and decision services
(5) User interfaces and external system integration.

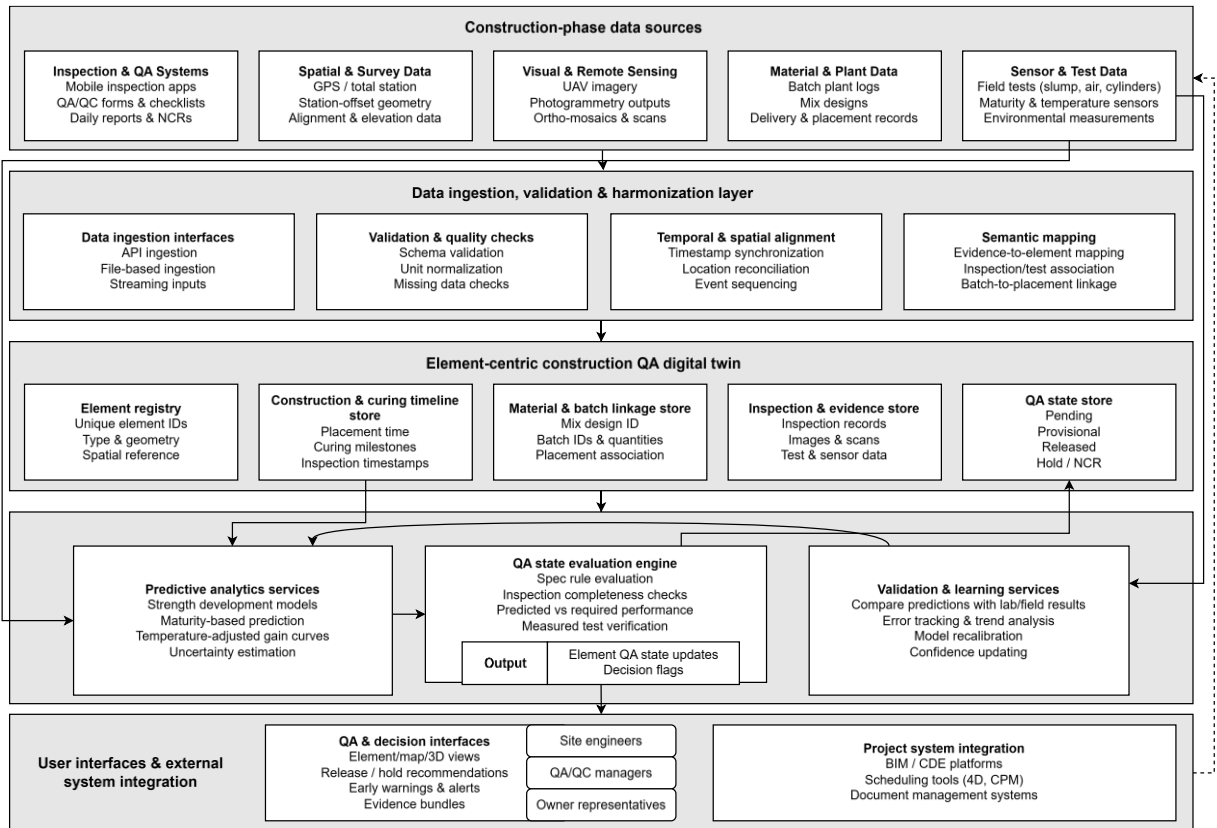

Fig. 2. Layered system architecture of the construction-phase QA digital twin.

The design separates data acquisition, data management, analytics, and decision logic to support modular implementation and future expansion, while maintaining traceability from raw data to final QA decisions.

### 3.1.1 Construction-Phase Data Sources

The architecture combines several types of construction phase data that are usually handled in an isolated manner. They are inspection and QA systems (mobile inspection applications, checklists, daily reports, and non-conformance records), spatial and survey data (GPS and total station measures, station-offset geometry and alignment and elevation data), visual and remote sensing data (UAV images, photogrammetric products and gmail), material and plant data (batch plant logs, mix designs and delivery and placement records), sensor and test data (field test results, maturity and temperature sensor readings and environmental measurements). Each data source is different in terms of format, resolution, temporal granularity and reliability. Instead of the system presupposing standardized inputs, the system architecture clearly supports this heterogeneity by downstream ingestion, validation and alignment processes.

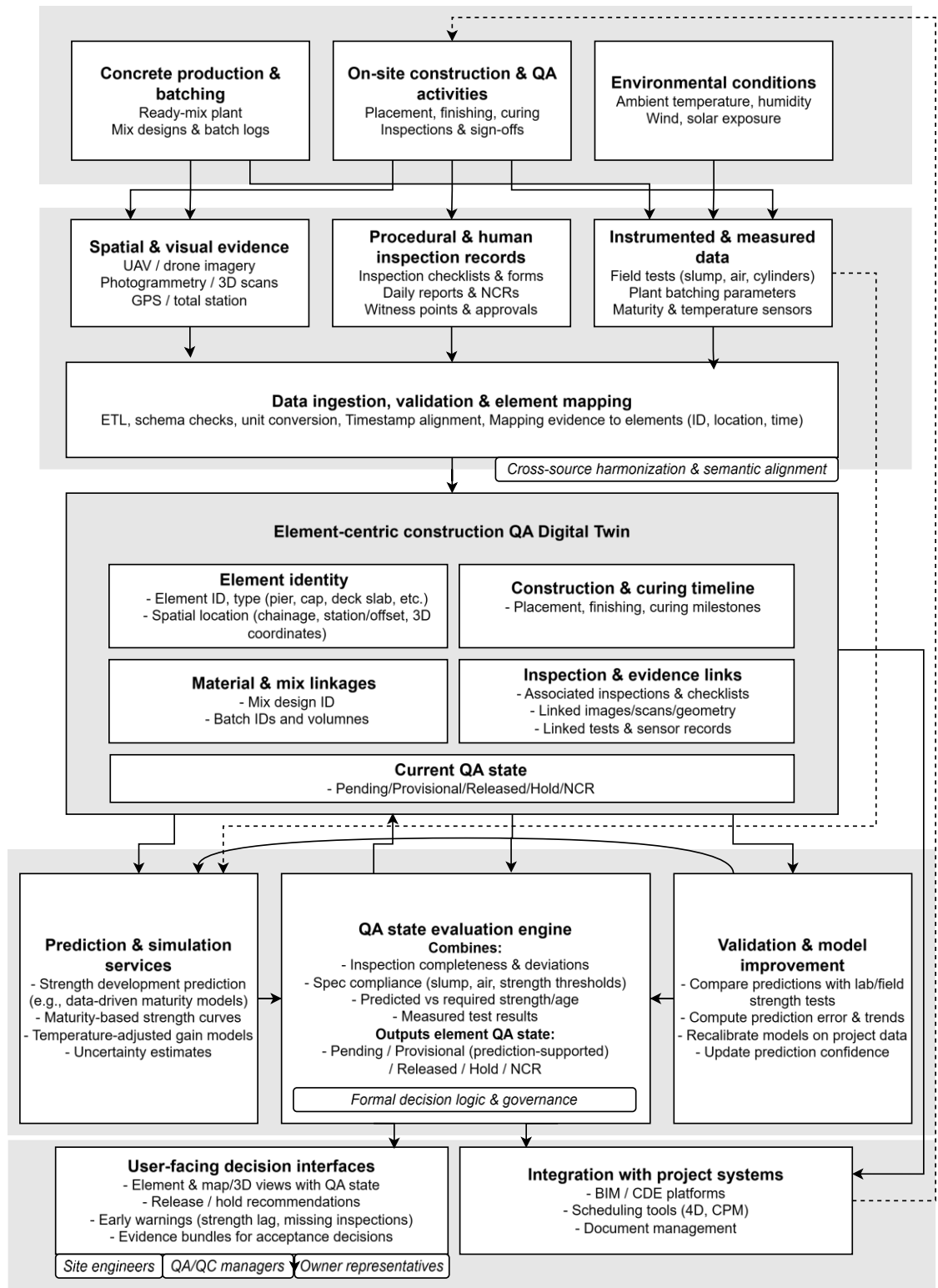

Fig. 3 . Operational data-to-decision workflow within the element-centric QA digital twin.

### 3.1.2 Data Ingestion, Validation and Harmonization Layer

The layer of data ingestion, validation, and harmonization is the technical base of the transformation of the heterogeneous data of construction to a coherent and analyzable form. This layer has various ingestion interfaces such as API-based integration, file-based ingestion, and streaming inputs which enable the system to connect with both the new digital tools and the old construction processes. The incoming data is subjected to validation and quality checks such as schema validation, unit normalization, and missing or inconsistent values. Processes of temporal and spatial alignment are used to synchronize the timestamps, match the location references, and sequence the construction events to allow inspection records, sensor data, and material records to be read in a shared spatiotemporal frame. Finally, semantic mapping operations are used to associate individual data records with construction elements so inspection observations, test results, and batch records can be associated with specific elements on the basis of location, time, and construction context. This mapping step is very important for facilitating element-level reasoning in succeeding layers.

### 3.1.3 The element-centric QA digital twin core

The element-centric construction QA digital twin is at the heart of the system architecture and will be the system of record of quality information at the construction phase. The digital twin does not organize data by document or activity, but by individual structural elements, which allows the assessment of QA to be fine-grained and traceable. The core data model consists of a number of inter-related elements. A construction and curing timeline store records the times when it was placed, when the material has cured, and when it was inspected, and offers some temporal information on the development of strength and the sequence of inspections. The material and batch linkage stores identify the elements with mix designs, batch ID, amounts, and placement documents. Inspection and evidence stores to maintain structured records of inspection, images, scans and related test and sensor data. In addition, a dedicated QA state store is used to maintain the current QA status of each element, including pending, provisional, released, hold, or non-conformance states. Such a clear description of the QA state allows systematic monitoring of preparedness and adherence during the construction process.

### 3.1.4 Analytics, Rule Evaluation, and Learning Services

There is an analytics and decision services that lie above the core digital twin to transform stored data into actionable QA insights. Predictive analytics services approximate the strength development based on maturity-based and temperature-adjusted models with uncertainty estimation to make decisions with incomplete information.

A central QA state evaluation engine is used to control quality decisions by integrating predictive outputs, checks completeness, specification rule evaluation, and measured test verification. This engine is responsible for evaluating whether design and specification requirements are met for each element and generates element-level QA state updates, and decision flags. Importantly, predictive results are considered as decision support, not acceptance authority, which ensures that they are in accord with contractual QA practices.

Validation and learning services close by the feedback loop comparing the predicted strength development with measured laboratory and field test results. Prediction errors and trends are monitored over time which allows model recalibration and updating of confidence on a project-specific basis. This closed-loop learning mechanism enables predictive services to evolve in accordance with local materials, environmental conditions and construction practices throughout the life of the project.

### 3.1.5 *User Interfaces and External System Integration*

The last architecture level exposes the outputs of the digital twin via user-interfacing and connecting with other project systems. QA and decision interfaces visualize element and map-based views of the QA state with recommendations of release and hold, early warnings of strength lag or missing inspections and evidence bundles to support acceptance decisions. These interfaces are intended to be used by site engineers, QA/QC managers and owner representatives (both operational and oversight).

In parallel, integration services enable the digital twin to be integrated with external systems of the project, such as BIM and common data environment (CDE) platforms, scheduling tools and document management systems. These integrations allow for QA state information to support the construction planning, sequencing and documentation without the duplication of data across systems.

The proposed system architecture is designed to be modular and independent of technology. By decoupling data ingestion, core data management, analytics and interfaces, the architecture can be implemented with different technologies with the same conceptual structure. Additional data sources, predictive models, or decision rules can be added without re-organizing the underlying digital twin, and this makes it easily extendable to any type of project and organization.

## 4   Element-Level QA Workflow Demonstration for Bridge Construction

To demonstrate how the proposed QA digital twin supports day-to-day construction decisions, a representative bridge workflow is considered: drilled shaft, column, cap, girder or deck panel installation, and deck steel and concrete placement (Fig. 4). Each stage is treated as a separate construction element with a defined QA gate. The next activity cannot proceed until the required inspections and material checks for that element are completed. For the drilled shaft, inspection records such as excavation conditions, reinforcement cage placement, slurry or cleanliness checks, and embedment details are collected. Related material testing records are also included. These records are ingested, validated, and linked to the specific element using its ID and location reference. The QA engine then evaluates whether inspection requirements and material criteria are satisfied. Based on this evaluation, the element is either released for the next activity (column construction) or placed on hold for corrective action.

The same process applies to the column and cap. Pre-placement inspections such as reinforcement layout, clear cover, formwork stability, and alignment are recorded. Placement information, including batch tickets, placement time, and ambient conditions, is also linked to the element. Field and laboratory test samples are associated in the same way. The QA state engine continuously evaluates completion and compliance before allowing the next construction stage to proceed. During girder and deck panel installation, the inspection focus shifts to geometry and fit-up checks, such as bearing seat elevations, line and grade, and panel alignment. At this stage, the

QA state also considers readiness constraints to prevent premature deck operations.

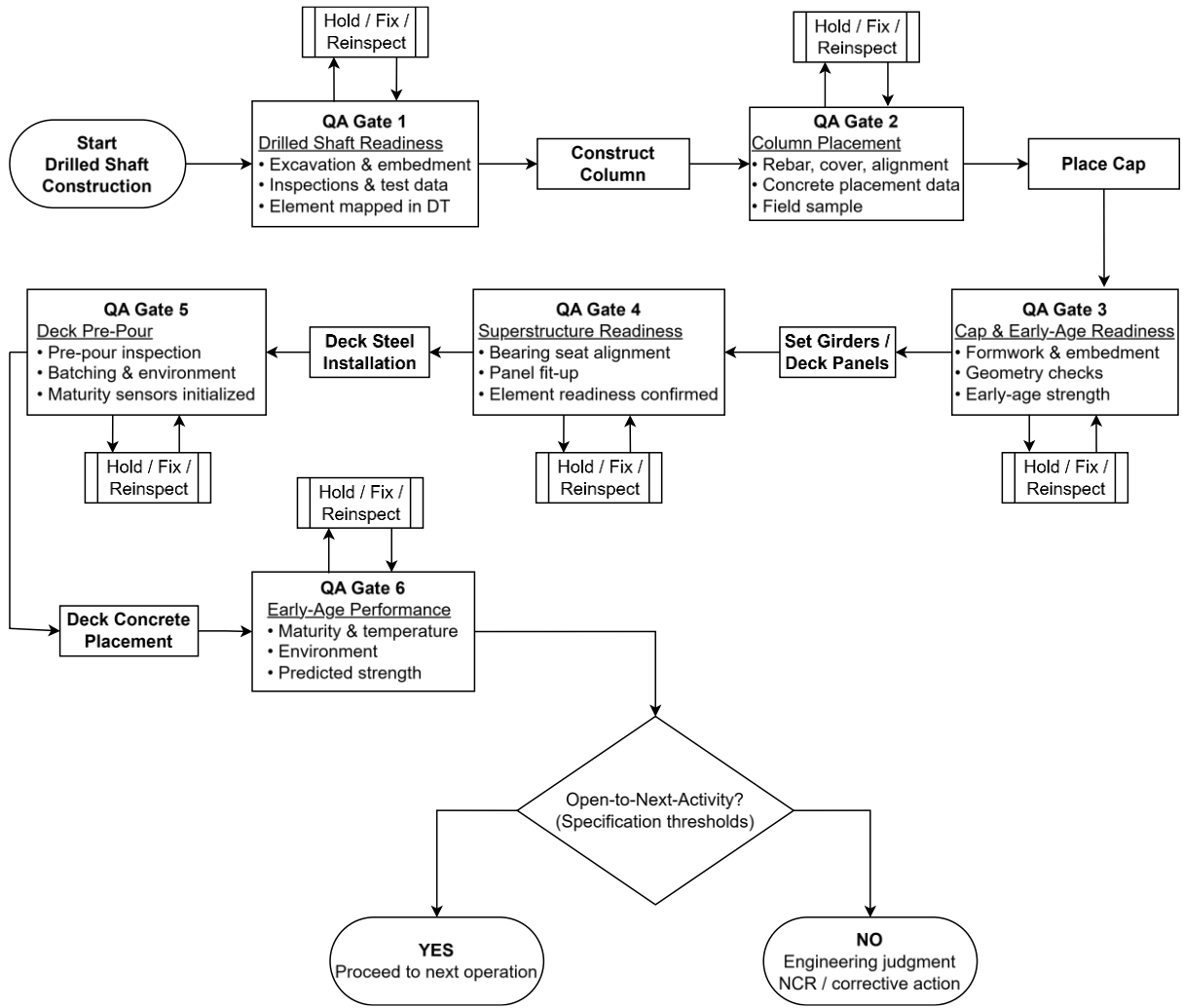

Fig. 4. QA/QC workflow for bridge construction activities.

Before deck steel placement and concrete pouring, pre-pour inspections are combined with material-side inputs such as batching parameters, environmental conditions, and maturity sensor initialization. The system uses this information to estimate early-age strength and assess whether readiness thresholds are likely to be achieved before standard 28-day test results are available. As laboratory and field test results become available, they are used to validate and update the predictive estimates. The QA decision gate then generates a clear and traceable action: release, hold, or initiation of a non-conformance report if needed. This example shows how element-level mapping, early warnings, and defined QA gates can help identify potential non-compliance earlier and reduce the risk of rework and schedule disruption.

# 5 Implementation Challenges and Practical Considerations

While the proposed construction-phase QA digital twin offers clear advantages, implementing it on large civil infrastructure projects is not straightforward. The challenges are often less about technology and more about contracts, data coordination, and field workflow.

## 5.1 Alignment with Specifications and Contracts

Quality acceptance in infrastructure projects is governed by agency specifications, contract documents, and project-specific provisions. These requirements may change during the project due to revisions, clarifications, or policy updates. Changes in testing frequency, acceptance limits, or documentation requirements directly affect how QA decisions must be made.

For this reason, the digital twin cannot rely on fixed rules. Its decision logic must remain configurable and traceable to current contract requirements. Continuous coordination between engineers and system administrators is necessary to ensure that the system reflects the latest approved specifications. Without this alignment, the digital twin could create confusion rather than clarity.

## 5.2 Data Integration Across Multiple Systems

Construction data are typically stored across different platforms. Inspection reports may be managed in one system, laboratory results in another, production data in a plant database, and geometry in BIM or survey tools. Each system may use different identifiers, naming conventions, and location references.

Linking these datasets at the element level is often difficult. Without consistent element identifiers or location standards, integration requires manual reconciliation. For a QA digital twin to function effectively, projects need agreed-upon naming conventions and reference systems early in the project lifecycle.

## 5.3 Integration with Batch Plant and Production Data

Material performance assessment depends on production data such as mix design parameters, batch times, and delivery records. However, batch plant systems are often proprietary and differ across vendors. Data formats, time resolution, and reporting structures are not always consistent.

These inconsistencies make it challenging to accurately match production data with placement records and inspection timelines. In some cases, batch information may be delayed or incomplete, which limits the reliability of early-stage performance assessments. The digital twin must therefore be able to operate under partial data conditions and clearly indicate associated uncertainty.

## 5.4 Timing and Data Latency

Construction QA data do not arrive at the same time. Laboratory test results may be available days or weeks after placement. Sensor data may be interrupted due to connectivity issues. Inspection records may be uploaded after field activities are completed.

These delays affect how the quality state of an element is interpreted at any given moment. The system must distinguish between provisional decisions based on early indicators and final decisions based on verified test results. Clear timestamps, revision history, and documentation of decision points are essential for maintaining transparency.

### *5.5   Workflow Integration and User Adoption*

Field inspectors and engineers already follow established reporting procedures. If a new system requires duplicate data entry or disrupts existing routines, adoption will be difficult.

Successful implementation depends on integrating the digital twin with current inspection tools and laboratory workflows rather than replacing them. The system should reduce coordination effort, not increase it. Demonstrating practical benefits—such as earlier identification of potential non-compliance or improved traceability—will be key to gaining user trust.

### *5.6   Governance and Accountability*

Quality acceptance decisions have contractual and legal implications. The digital twin must maintain clear accountability. Role-based access control, documented approvals, and a transparent audit trail are necessary to ensure that decisions remain defensible.

Importantly, the system should support engineering judgment rather than replace it. Final acceptance authority must remain with designated responsible parties, consistent with contract requirements.

## 6   Conclusion and Future Work

This paper presented a construction-phase QA digital twin framework designed to support element-level decision making in large civil infrastructure projects. The framework integrates inspection records, material testing results, and predictive analytics into a unified system that reflects the evolving quality state of individual construction elements. By linking inspection findings, production and placement data, early-age material behavior, and specification-based acceptance logic, the approach addresses limitations of traditional QA processes that rely on delayed test results and fragmented documentation.

A key contribution of this work is treating quality assurance as a time-dependent process rather than a final evaluation after construction activities are completed. The proposed framework enables earlier identification of potential non-compliance, supports readiness-based progression to subsequent construction stages, and maintains traceability between inspection evidence, test data, and acceptance decisions. The system architecture and bridge workflow example demonstrate that the framework can complement existing inspection and testing procedures without replacing them, while providing structured integration and decision support.

Future work will focus on broader field implementation across different project types and material systems, including cement-treated base and asphalt mixtures. Additional research is needed to develop standardized data schemas and element identification strategies to improve interoperability across owner, designer, contractor, and supplier platforms. Project-specific calibration and uncertainty evaluation of predictive models will also be important to strengthen confidence in early-age decision support. Finally, alignment with agency QA policies and digital delivery standards will be necessary to support wider adoption.

In general, the suggested construction-stage QA digital twin offers a feasible route to enhancing quality control measures practices by moving focus toward a more data-driven, proactive decision making at the construction stage.